\documentclass[twocolumn,superscriptaddress,floatfix,preprintnumbers,showpacs,amsmath,amssymb]{revtex4-1}

\usepackage{amsmath}
\usepackage{natbib}
\usepackage{amssymb}
\usepackage[dvips]{color}
\usepackage{graphicx}
\usepackage{dcolumn} 
\usepackage{bm} 
\usepackage{hyperref}
\usepackage{colortbl}

\begin{document}

\title{Observation of Wannier-Stark localization at the surface of BaTiO$_3$ films by photoemission}

\author{Stefan~Muff$^{1,2}$}
\author{Nicolas Pilet$^{2}$}
\author{Mauro~Fanciulli$^{1,2}$}
\author{Andrew~P.~Weber$^{1,2}$}
\author{Christian Wessler$^{2}$}
\author{Zoran~Risti\'{c}$^{2}$}
\author{Zhiming~Wang$^{2,3}$}
\author{Nicholas~C.~Plumb$^{2}$}
\author{Milan~Radovi\'{c}$^{2,4}$}
\author{J.~Hugo~Dil$^{1,2}$}

\affiliation{
$^{1}$ Institut de Physique, \'{E}cole Polytechnique F\'{e}d\'{e}rale de Lausanne, CH-1015 Lausanne, Switzerland
\\ 
$^{2}$ Swiss Light Source, Paul Scherrer Institut, CH-5232 Villigen, Switzerland
\\
$^{3}$Department of Quantum Matter Physics, University of Geneva, CH-1211 Geneva 4, Switzerland
\\
$^{4}$ SwissFel, Paul Scherrer Institut, CH-5232 Villigen, Switzerland
}

\date{\today}

\begin{abstract}
Observation of Bloch oscillations and Wannier-Stark localization of charge carriers is typically impossible in single-crystals, because an electric field higher than the breakdown voltage is required. In BaTiO$_3$ however, high intrinsic electric fields are present due to its ferroelectric properties. With angle-resolved photoemission we directly probe the Wannier-Stark localized surface states of the BaTiO$_3$ film-vacuum interface and show that this effect extends to thin SrTiO$_3$ overlayers. The electrons are found to be localized along the in-plane polarization direction of the BaTiO$_3$ film. 

\end{abstract}


\maketitle

Electric fields are the driving force of electric transport and a variety of electronic properties of semiconductor systems. The ultimate limiting mechanism of conductance in crystals is defect scattering, which prevents ballistic transport. However, in an ideal system, without defect scattering, electrons would perform an oscillating motion for large enough electric fields. 
These so-called Bloch oscillations form due to the Bragg scattering of the accelerated electrons at the Brillouin zone boundary \cite{Bloch:1929, Zener:1934}. The oscillations eventually lead to a Wannier-Stark localization (WSL) of the accelerated electrons as well as the formation of a Wannier-Stark ladder \cite{Marder:2010, Hofmann:2014}. 
These effects set a fundamental limit to coherent transport in crystals and their existence in a real system will provide further insight into its transport mechanisms.

The Bloch oscillation time for one cycle is given by $\tau_B=h/(eFa)$, where $h$ is the Planck constant, $e$ the electron charge, $F$ the electric field present  and $a$ the lattice parameter in the direction of the electric field. In real crystals, the critical condition for Bloch oscillations to be possible is a $\tau_B$ smaller than the relaxation time $\tau$ of the system which is determined by the mean free path $\lambda$ and the Fermi velocity $v_F$. In other words the electron has to complete one (or several) periods of the Bloch oscillation before being scattered at random lattice defects. This condition can not be met by applying an external electric field on single crystal semiconductors, because the required fields are orders of magnitudes higher than the breakdown voltages of these systems.
This problem was successfully addressed by the engineering of artificial semiconductor lattices of high quality. In artificial superstructures, the lattice parameter $a$ is increased and thus the required electric field is lowered to an achievable value to observe Bloch oscillations and related effects \cite{Esaki:1970, Mendez:1988, voisin:1988, vonPlessen:1992, Feldmann:1992, Waschke:1993}. Alternatively, Bloch oscillations can also be induced for a short time scale using terahertz radiation \cite{Schubert:2014}.

In this Letter we take a new approach, utilizing the electric field present in ferroelectric BaTiO$_3$ (BTO) to directly observe WSL of the two-dimensional states present at its surface by angle-resolved photoelectron spectroscopy (ARPES). Furthermore, it will be shown that this effect can be extended to thin overlayers of SrTiO$_3$ (STO) where the ferroelectric field of the BTO substrate localizes the STO surface states.

BTO is a well known ferroelectric material and closely related to the perovskites SrTiO$_3$, CaTiO$_3$ (CTO) and KTaO$_3$ (KTO). These materials are all known to host a two-dimensional electron gas at their surface \cite{Santander:2011, Meevasana:2011, King:2012, Santander:2012, Plumb:2014, Rodel:2016, Muff:2017}. Ferroelectric properties due to local lattice relaxations play, together with oxygen vacancies, a key role in the formation of the two-dimensional electron gas on the surfaces of these systems. These three perovskites are all classified as incipient ferroelectrics in which quantum fluctuations prevent a ferroelectric order \cite{Weaver:1959, Muller:1979, Zhong:1996, Lemanov:1999}. Two-dimensional states can also be expected at the surface of BTO, which presents an excellent opportunity to study the impact of bulk ferroelectric order on these two-dimensional states at the surface and their transport properties \cite{urakami:2007}. It also offers a means of inducing ferroelectricity in other perovskites through doping or multilayer structure assembly. 
With the insights into fundamental transport mechanisms by the observation of WSL, new ways to directly manipulate and tailor transport properties of ferroelectric semiconductors become accessible.

\begin{figure}[tb]
	\includegraphics[width=0.5\textwidth]{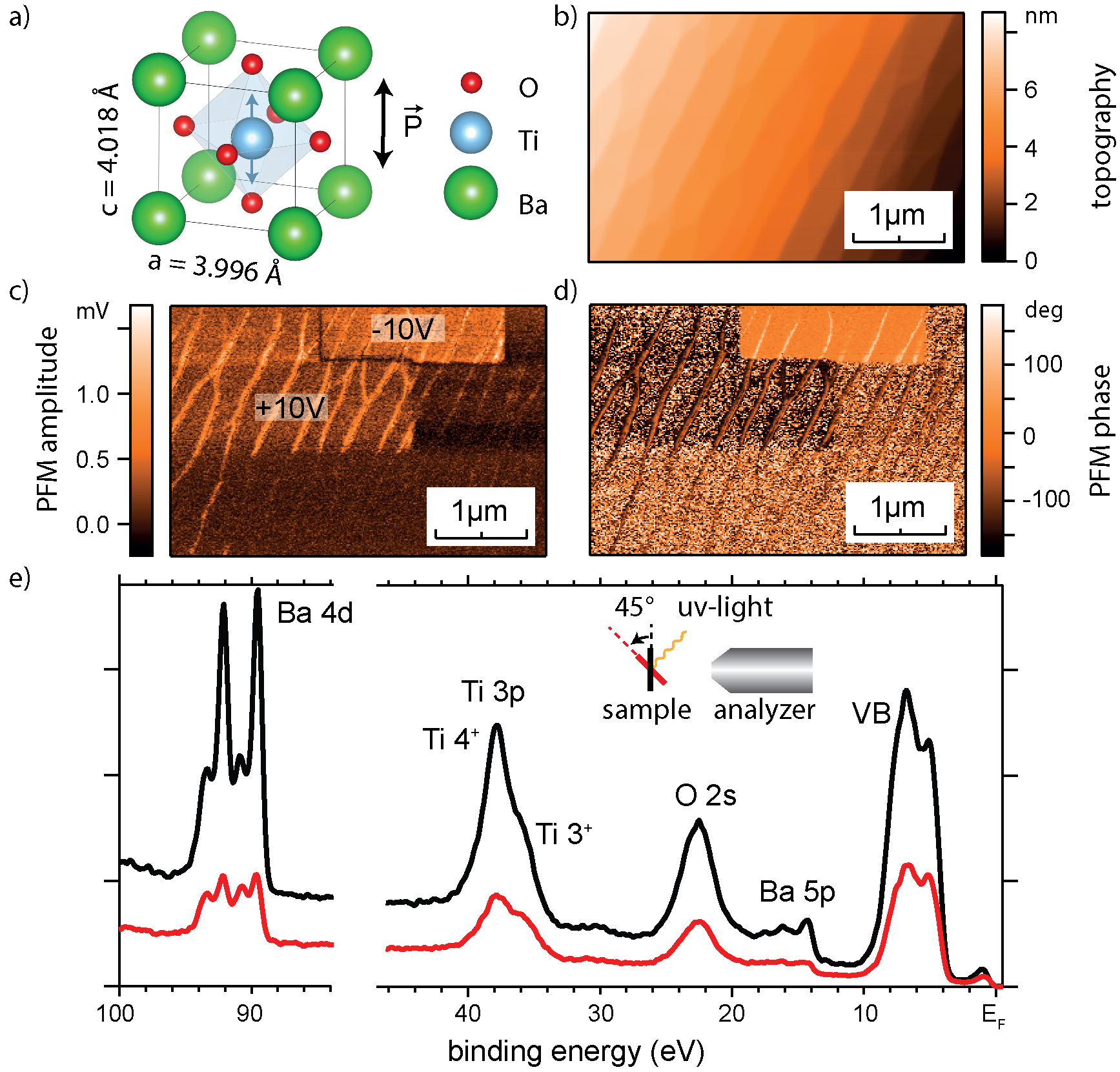}
	\caption{(a) Tetragonal BTO uc with indicated polarization axis. (b) AFM topography of a film of 50 uc BTO grown on Nb:STO. (c) and (d) PFM phase and amplitude. Square on the left side is poled by +10V applied to the probe tip, the square on the top with -10V as indicated. (e) XPS measurements with $h\nu=170$~eV photons for normal emission (black) and an emission angle of 45$^{\circ}$ (red).}
	\label{fig:pfm_xps}
\end{figure}

Bulk crystalline BTO is ferroelectric below the transition temperature of 120$^{\circ}$, exhibiting three different ferroelectric phases \cite{Merz:1949, Potnis:2011}.
The phase diagram of BTO thin films is significantly different compared to bulk BTO.
In films, the transition temperature is raised for compressive as well as tensile strain \cite{Li:2006}. For tensile strain, it has been demonstrated that solely an orthorhombic phase exists below the ferroelectric transition temperature \cite{Tenne:2004,Li:2006,Dionot:2014}. For compressive strain the tetragonal phase [Fig.~\ref{fig:pfm_xps}(a)] is the only ferroelectric phase present below the transition temperature, with a preferred polarization along the out-of-plane axis in films with a thickness of around of 5 unit cells (uc) \cite{Paul:2007,Dionot:2014}.
With increasing film thickness, strain and growth-defect relaxation will be responsible for a mixture of in-plane and out-of-plane domains of the tetragonal phase. In thicker films of BTO grown on STO, a mixture of domains with the size of around 20~nm can be observed as a result of relaxation \cite{Li:2006, Dubourdieu:2013}. Furthermore, the formation of domain walls with a 90$^{\circ}$ change in polarization direction are preferred energetically to 180$^{\circ}$ domain walls \cite{Dionot:2015}. This also favors a mixture of domains with out-of- and in-plane polarization directions.

The films investigated in this work were grown by pulsed laser deposition (PLD), allowing a controlled layer-by-layer growth monitored by reflective high-energy electron diffraction (RHEED). Films with a thickness of 20~uc where grown on commercially available, single-terminated SrTiO$_3$, Nb:SrTiO$_3$ and KTaO$_3$ (001) substrates (SurfaceNet GmbH, see \cite{SOM:BTO}). The growth was performed at a substrate temperature of 950~K, in a partial oxygen pressure of $1\cdot10^{-5}$~mbar. STO films of 3 and 5~uc thickness where grown on top of this BTO film under similar conditions (see \cite{SOM:BTO}).
The samples were \textit{in-situ} transferred to the high-resolution ARPES endstation and measured with circularly polarized synchrotron light. During the measurements the sample was held at 20~K and kept in ultra high vacuum (UHV) conditions better than $1\cdot 10^{-10}$~mbar.
The films were \textit{ex-situ} transferred to the NanoXAS beamline for piezo-response force microscopy (PFM) measurements at room temperature under UHV condition. The sample measured with PFM had a thickness of 50~uc and was grown on a 0.5wt\% Nb doped STO substrate under the same conditions as described above. A conductive substrate was chosen in order to have a well-defined back electrode and the higher film thickness in favor of a stronger PFM response signal.  

The PFM topography in Fig.~\ref{fig:pfm_xps}(b) shows a uniform sample surface \cite{SOM:BTO} according to the vicinality of the substrate, where a 0.2$^{\circ}$ miscut to the (001) surface was chosen to promote a layer-by-layer growth.  The PFM phase and amplitude in Fig.~\ref{fig:pfm_xps}(c,d) of the as-grown sample (bottom part of the field of view) shows no noticeable contrast, indicating that no intrinsic domains of resolvable size ($\gtrsim 20$~nm) are present. After subsequent writing of the surface with +10~V and -10~V applied to the probe tip, a phase and amplitude contrast is noticeable, proving the presence of ferroelectric properties in our films. However, both of the written regions exhibit significant noise, reducing the difference of the mean phase value between the positive and negative poled region to approximately $72^{\circ}$ \cite{SOM:BTO}. This difference is significantly less than the $180^{\circ}$ phase difference expected for completely opposite polarized regions. The reasons for this observation is a not completely homogeneously polarized surface in the written areas. This indicates a strong locking of the domains in the in-plane direction due to interface strain and relaxation mechanisms. 

\begin{figure*}[htb]
	\includegraphics[width=1\textwidth]{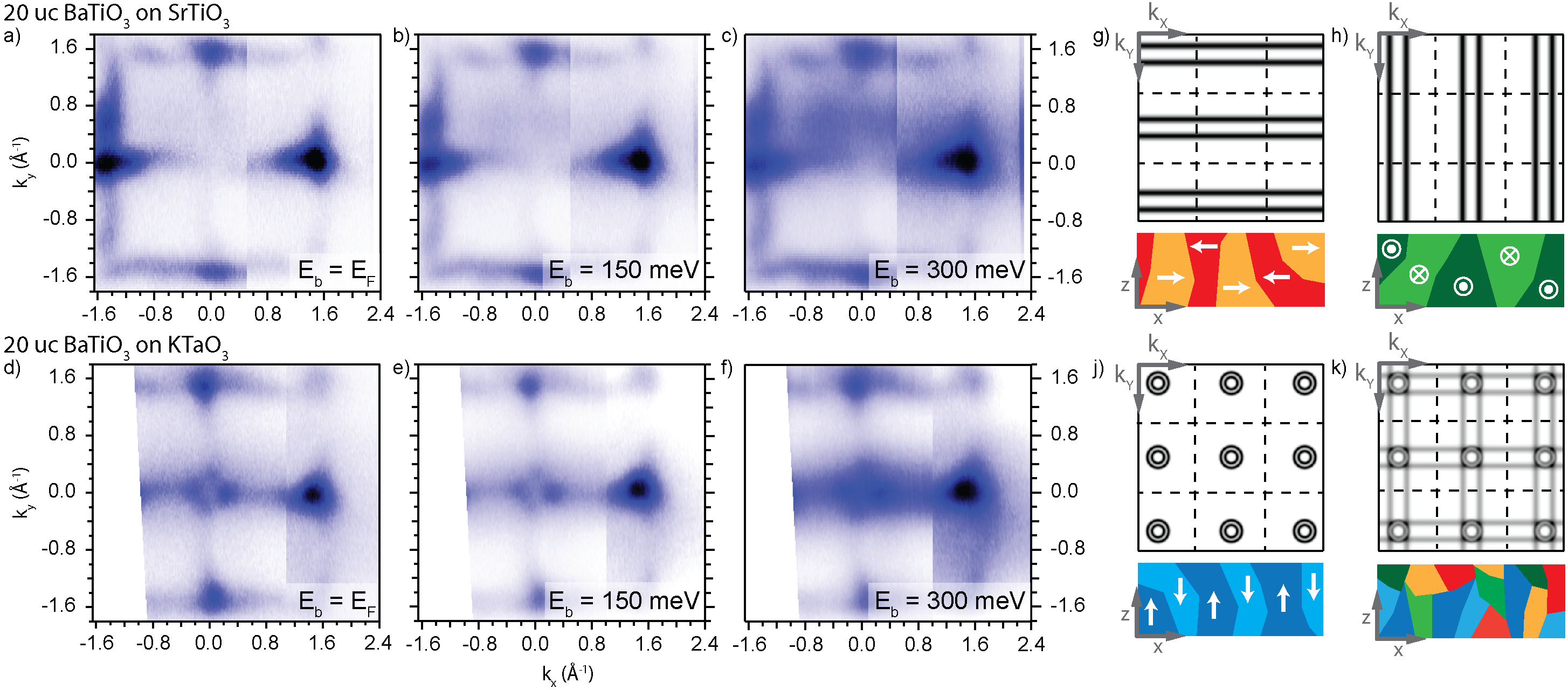}
	\caption{(a-c) Constant energy surfaces of 20 uc of BTO grown on STO measured with $h\nu=80$~eV for $E_b=E_F$ (a), $E_b=150$~meV (b) and $E_b=300$~meV (c). (d-f) Same as (a-c) for 20 uc BTO on KTO. (g-k) Different ferroelectric domain configurations of the films and the corresponding WSL states for in-plane polarizations along $\left\langle 100 \right\rangle$ (g) and $\left\langle 010 \right\rangle$ (h) and out-of-plane polarization along $\left\langle 001 \right\rangle$ (j). (k) Combined WSL states from the three configurations with equal weight.}
	\label{fig:arpes80eV_model}
\end{figure*}

The X-ray photoelectron spectroscopy (XPS) measured at normal emission and at a more surface sensitive emission angle of 45$^{\circ}$ show the Ba 4d, Ti 3p and O 2s core levels [Fig.~\ref{fig:pfm_xps}(e)]. The Ba 4d core level consists of the spin-orbit split Ba 4d$_{3/2}$ and 4d$_{5/2}$ doublet and a lower intensity doublet, shifted by 1.25~eV to higher binding energies. By comparing the peak areas of the two species for the two emission angles we can assign the higher binding energy, chemically-shifted doublet to undercoordinated Ba ions in the BTO surface region \cite{Jacobi:1987, Hudson:1993, Cai:2007}. The Ti 3p core level includes two peaks assigned to the Ti 4$^+$ and Ti 3$^+$ ions whereby the latter is more surface localized. Close to the Fermi energy an in-gap state is located at a binding energy of $0.8$~eV in the bulk band gap \cite{Rault:2013, SOM:BTO}. The intensity of the in-gap state and the Ti $3^+$ peak increases under UV irradiation \cite{SOM:BTO}. Similar effects are observed in the case of STO where the presence of Ti 3$^+$ ions and the in-gap state are explained by surface relaxation and oxygen vacancies \cite{Plumb:2014}. Comparing the relative Ba 4d and Ti 3p peak areas for the two emission angles, we can conclude that the surface is TiO$_2$ terminated \cite{Radovic:2009}. 

\begin{figure*}[htb]
	\includegraphics[width=1\textwidth]{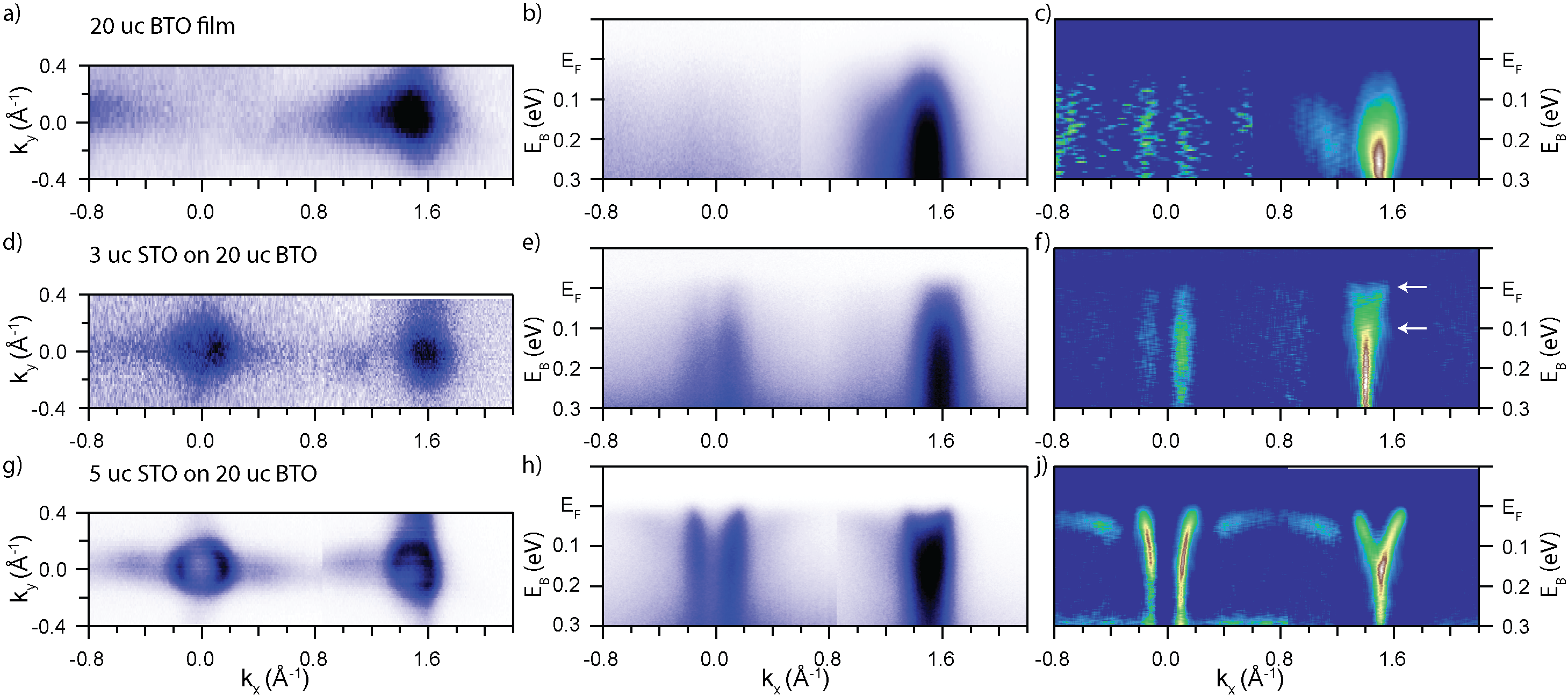}
	\caption{(a,d,g) Fermi surface, (b,e,h) band map at $k_y=0$~\AA$^{-1}$, and (c,f,j) two-dimensional curvature \cite{Zhang:2011curv} measured at $h\nu=82$~eV. (a-c) For the 20~uc BTO film, (d-f) for a 3~uc STO film on top of 20~uc of BTO, and (g-j) for a 5~uc STO film on top of 20~uc of BTO. The arrows in (f) indicate the polaron replica.}
	\label{fig:STOonBTO}
\end{figure*}

The ARPES measurements in Fig.~\ref{fig:arpes80eV_model} and Fig.~\ref{fig:STOonBTO}(a-c) show metallic states emerging from the in-gap state. Resonant effects cause strong intensity modulations, but the states show no clear dispersion as a function of photon energy, indicating their two-dimensional nature \cite{SOM:BTO}.
As for STO, matrix element effects are responsible for the suppression of intensity at $k_x=0$~\AA$^{-1}$ due to the mainly xy-symmetry of the two-dimensional state \cite{SOM:BTO}. These states can be attributed to the partially filled Ti 3$d_{xy}$ orbital, that is split from the Ti 3$d_{xz}$ and Ti 3$d_{yz}$ orbitals due to a distortion of the TiO$_6$ octahedron by lattice relaxation \cite{Santander:2011, Plumb:2014, Muff:2017}. 
However, the Fermi surface around the $\overline{\Gamma}$-points shows no clear bands of the two-dimensional states but features spectral weight, elongated along both $\overline{\Gamma \mbox{X}}$-directions.  
These elongated states extend over multiple surface Brillouin zones connecting the neighboring $\Gamma$-points as shown in Fig.~\ref{fig:arpes80eV_model}(a,d). Comparing the Fermi surface with constant energy surfaces at higher binding energies [Fig.~\ref{fig:arpes80eV_model}(b,c) and (e,f)], no dispersion of these states with respect to the binding energy is noticeable. On the other hand, at the binding energy of the in-gap state the spectral weight is smeared out equally in all directions and no pattern is discernible \cite{SOM:BTO}. As we will explain below, the absence of dispersion in the states around the Fermi level is due to WSL and a direct consequence of the electric field present in the bulk of the film.

The electrons in the two-dimensional state experience an accelerating force in the direction opposite to the electric field present in the ferroelectric domains. Due to the potential barrier at unit cell boundaries, the acceleration is not uniform but is described by Bloch oscillations \cite{Bloch:1929, Zener:1934}. This localizes the electron in real space and hence shows smearing in reciprocal space. 
Considering the lattice parameter of BTO, the condition $\tau>\tau_B$ for Bloch oscillations to exist is fulfilled for an electric field $F \gtrsim 10^9$~V/m assuming a typical relaxation time of $\tau = \lambda / v_F \approx 10^{-14}$~s \cite{Hofmann:2014}, which exceeds the breakdown field strength of known insulators. 
Due to its ferroelectric properties, the local electric field in the BTO film is several order of magnitudes higher than any possible external electric field. 
An estimate of the electric field inside BTO films can be obtained from the spontaneous polarization which is $P\approx0.25~C/m^2$ for tetragonal bulk BTO \cite{Merz:1953} and is predicted to increase for strained films \cite{ederer:2005}. The resulting local electric field of $1\cdot10^{10}$~V/m in ferroelectric BTO is in the order of the electric field required to meet the condition for Bloch oscillations \cite{SOM:BTO}. Bloch oscillations are typically accompanied by the formation of a Wannier-Stark ladder, a set of electronic states separated in energy and space \cite{SOM:BTO}. Due to the surface localization of the Bloch oscillations in the BTO films, the observation of the WS ladder is beyond the compatibility of conventional techniques \cite{Esaki:1970, Mendez:1988, voisin:1988, vonPlessen:1992, Feldmann:1992, Waschke:1993}. For such systems the WSL induced smearing observed by ARPES provides an alternative method to observe these effects.  

The BTO films grown on a STO substrate have a compressive strain of 2~\% at the interface. As a result, the film is expected to stay in a single tetragonal phase below the ferroelectric phase transition \cite{Paul:2007, Dionot:2014}. Tetragonal BTO can host a ferroelectric polarization along the $\left\langle 001 \right\rangle$ out-of-plane as well as the $\left\langle 100 \right\rangle$ and $\left\langle 010 \right\rangle$ in-plane directions. While at the interface the polarization direction is preferably along the out-of-plane axis, strain relaxation mediated by growth defects will be responsible for a mixture of domains close to the film surface. 
The domains with different electric field directions will all contribute differently to the Fermi surface. The in-plane domains, exhibiting an electric field along $\left\langle 100 \right\rangle$ [Fig.~\ref{fig:arpes80eV_model}(g)] or $\left\langle 010 \right\rangle$ [Fig.~\ref{fig:arpes80eV_model}(h)] directions will both give rise to WSL. In ARPES, this WSL becomes visible as one-dimensional states along the $\overline{\Gamma \mbox{X}}$-directions. In domains where the electric field is along the out-of-plane or $\left\langle 001 \right\rangle$ directions [Fig.~\ref{fig:arpes80eV_model}(j)], the electric field will lift the spin-degeneracy of the two-dimensional states. The resulting Rashba-type spin splitting consists of oppositely spin-polarized, concentric rings at the Fermi surface \cite{Dil:2009R,Santander:2014, KrempaskyPRB:2016}. The direction of the spin polarization of the bands will be inverted depending on the sign of the ferroelectric polarization vector.

With a domain size on the order of 20~nm \cite{Dubourdieu:2013}, the synchrotron beam with a size of around 100~$\mu$m will average over several domains with different ferroelectric polarization directions. The resulting model Fermi surface in Fig.~\ref{fig:arpes80eV_model}(k), formed by an overlay of the contributions from the different domains is in good agreement with the ARPES measurements, especially if further modulation of the ARPES signal by matrix element effects are considered.  

For BTO films grown on KTO the compressive strain is reduced to 0.2~\% due to the larger lattice constant of KTO compared to STO. With the change in strain also the domain formation is expected to be different for the BTO films on KTO. Furthermore, our KTO substrates have a higher step density as our STO substrates inducing an imbalance between different domains \cite{SOM:BTO}. When comparing the data of the BTO film on KTO [Fig.~\ref{fig:arpes80eV_model}(d-f)] with the results of the film grown on STO [Fig.~\ref{fig:arpes80eV_model}(a-c)] two differences are noticeable: i) the signature of a circular Fermi surface contribution around $\overline{\Gamma_{00}}$ and ii) the WSL states are more intense along the $k_x$- than the $k_y$-direction. Both of these observations are in agreement with an altered domain configuration. The reduced interface strain and the higher step density are responsible for the formation of larger domains with a higher fraction polarized along the z- and x-directions in the measured BTO films grown on KTO.

The general nature of the WSL is illustrated by its presence in ultrathin STO films grown on top of the BTO layers. The ARPES data for a 3~uc thick STO film in Fig.~\ref{fig:STOonBTO}(d-f) exhibits states very similar to pure BTO Fig.~\ref{fig:STOonBTO}(a-c). The Fermi surface shows stripes extending over several surface Brillouin zones (see \cite{SOM:BTO}) characteristic for WSL. However, the reduced electric field with increasing STO film thickness results in a lower intensity of the smearing and a shallow electron pocket with polaron replicas \cite{Moser:2013, Wang:2016} becomes visible [see markers in Fig.~\ref{fig:STOonBTO}(f)]. For the surface of the 5~uc film in Fig.~\ref{fig:STOonBTO}(g-j) the fields of the BTO substrate are so far reduced that no indication of WSL is visible. The ARPES data resembles the electronic structure of bulk STO with the more filled circular $d_{xy}$ states forming the 2DEG and elongated $d_{xz}$ and $d_{yz}$ states that clearly disperse with binding energy (Fig.\ref{fig:STOonBTO}(h,j)) \cite{Plumb:2014}.

From these results we can conclude that either up to 3~uc of STO on BTO are ferroelectric, or that about 4~uc of STO are needed to let the electric field of BTO decay to a value that no longer influences the electronic properties at the sample surface. Furthermore, the sharp electron-like states at the surface of the STO films verify the high crystalline quality also of our BTO layer. RHEED and XPS data (see \cite{SOM:BTO}) indicate a layer-by-layer growth of the STO, opening the possibility to study whether a ferroelectric order is induced in the STO by high resolution transition electron microscopy in future work. 

To conclude, we have presented combined effects of two different physical properties on BTO film surfaces: the formation of a two-dimensional state and the Wannier-Stark localization of these state. We have further demonstrated that ARPES provides a novel means of probing WSL in reciprocal space. To our knowledge, this is the first time WSL is directly observed in a single crystalline semiconductor. The combined presence of electric fields and two-dimensional states at the surface of a transition metal oxide opens up a rich field to study the interplay of ferroelectricity and interface states.  For the study of the macroscopic influence of WSL on the transport properties, BTO films with preferred polarization directions should be prepared by the help of different substrates regarding orientation, lattice parameters and conductivity \cite{Sinsheimer:2013, ChenJ:2013, Guo:2016} and under different growth conditions \cite{Rault:2013}.  
Furthermore, our experiments suggest that WSL should be a general effect for ferroelectric materials with surface or interface states, and domains with an in-plane electric field.

This work was financially supported by the Swiss National Science foundation (SNF) Project No. PP00P2\_144742/1 and Project No. 200021-159678.  

\bibliographystyle{apsrev4-1}

\newpage
\pagebreak
\onecolumngrid
\clearpage
\begin{center}
\textbf{\LARGE Supplemental Material \\ \qquad \\ \large Observation of Wannier-Stark localization at the surface of BaTiO$_3$ films by photoemission}
\end{center}
\setcounter{equation}{0}
\setcounter{figure}{0}
\setcounter{table}{0}
\setcounter{page}{1}
\makeatletter
\renewcommand{\thetable}{S\arabic{table}}  
\renewcommand{\thefigure}{S\arabic{figure}}
\renewcommand{\theequation}{S-\arabic{equation}}
\renewcommand{\bibnumfmt}[1]{[S#1]}
\renewcommand{\citenumfont}[1]{S#1}


\section{Piezo-Response Force Microscopy}

\begin{figure*}[htbp]
	\includegraphics[width=0.9\textwidth]{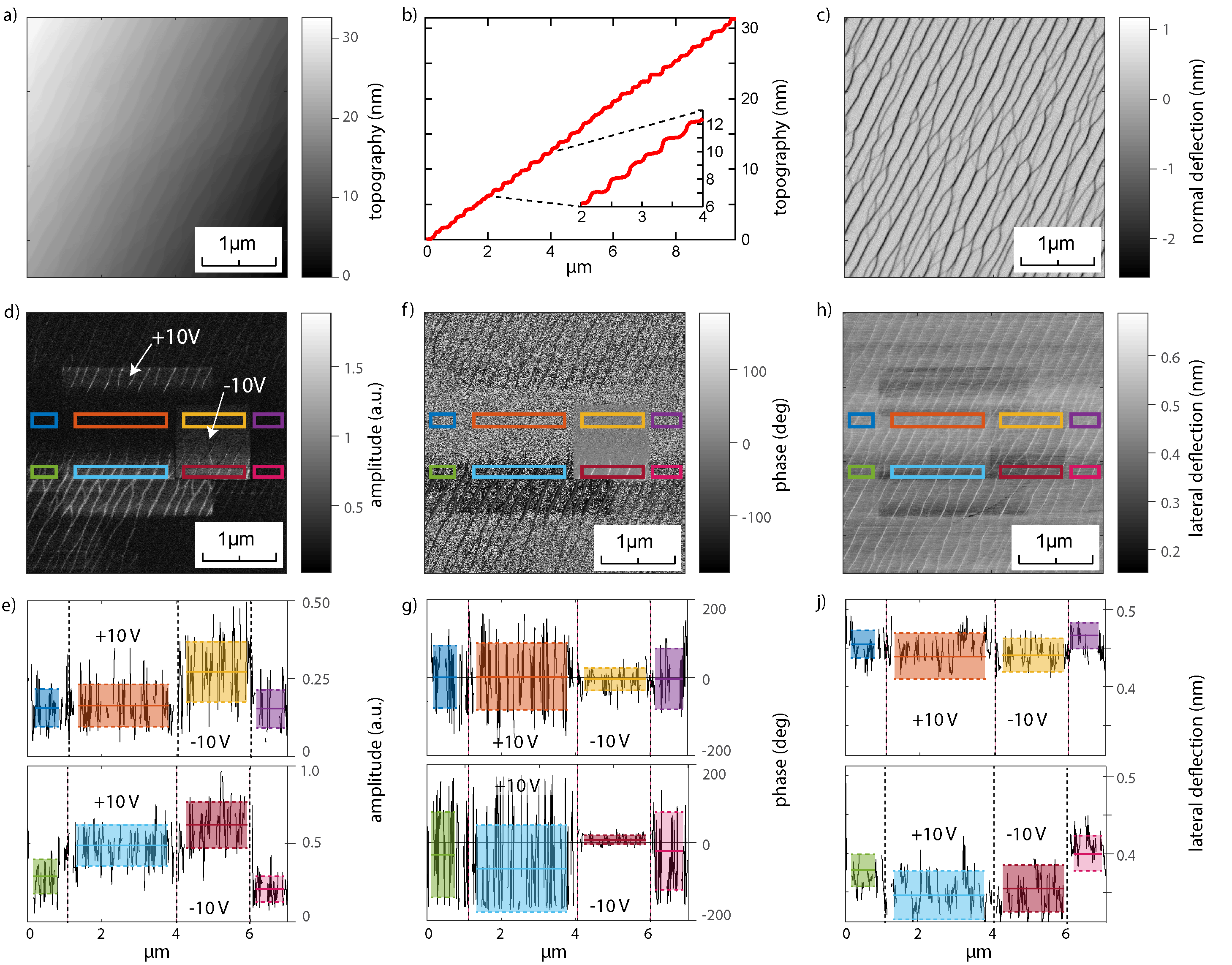}
	\caption{(a) PFM topography of 50 uc BTO grown on Nb:STO. (b) Topographic line profile along the diagonal perpendicular to the step edges. (c) Normal deflection. (d) PFM amplitude including written areas for +10~V and -10~V applied to the tip, as indicated. (e) Line profiles along the center of the marked squares in (d) with indicated average values and standard deviation of the marked areas. (f) PFM phase with (g) line profiles and (h) lateral deflection with (j) line profiles. }
	\label{fig:pfm}
\end{figure*}

The piezo-response force microscopy (PFM) data of Fig.~\ref{fig:pfm} is the full data set of the measurements presented in Fig.1 (b-d) of the main text. The topography [Fig.~\ref{fig:pfm}(a)] features a uniform step formation over the whole measured range with step heights [Fig.~\ref{fig:pfm}(b)] corresponding to one or multiple unit cells. The position of the step edges are visible in the normal deflection [Fig.~\ref{fig:pfm}(c)]. The terrace width of $0.1-0.2~\mu$m is given by the substrate misscut to the (001) surface of 0.1$^{\circ}$-0.2$^{\circ}$. 
The PFM amplitude and phase [Fig.~\ref{fig:pfm}(d,f)] show a visible contrast between the oppositely written areas as well as the unwritten area. The as-grown sample shows no formation of domains larger than the measurement lateral resolution of $\approx$20~nm. 
Due to the small PFM signal of the BaTiO$_3$ (BTO) film, the measurements where performed at the PFM resonance. This is causing a small cross talk of the PFM phase and amplitude with the topography signal, responsible for the visibility of the step edges in the PFM channels.

Two line profiles are taken along the scanning direction in the center of the marked, colored squares for the PFM amplitude [Fig.~\ref{fig:pfm}(e)] and phase [Fig.~\ref{fig:pfm}(g)] each. The signal in the vicinity of the step edges are excluded from the line profiles. 
In the amplitude line profiles a clear difference between the positively, the negatively, and the unwritten areas is noticeable. Especially the amplitude of the negative poled region is clearly higher than the positive and the unwritten areas. On the phase signal a clear reduction of noise is noticeable for the negatively poled area, while for the positively poled part the noise is on the level of the unpoled region (see Table \ref{tab:pfm}). The phase difference between the oppositely poled area is 6$^{\circ}$ for the top row and 72$^{\circ}$ for the bottom row with an error margin larger than the difference.

The measurements show that writing with a negative potential applied to the probe tip has a more noticeable effect than writing with a positive potential. Especially the strong noise reduction for the phase and the offset in the amplitude is obvious. In the phase, the mean value of the negative written area is, as for the unwritten part, very close to 0$^{\circ}$. This indicates that a larger portion of the intrinsic domains are polarized pointing out of the plane than into the plane. Thus the writing with positive potential applied is less effective. In order to uniformly polarize the two areas by switching all the in-plane domains to the out-of-plane axis, a higher potential than 10~V would be needed. 
In the lateral deflection [Fig.~\ref{fig:pfm}(g,h)] of the probe tips a clear contrast is noticeable between the written regions and the unpoled area. This lowering of the friction at the previously written areas could be due to the reduction of in-plane domains in these regions changing the local polarization fields.

\begin{table}[htbp]
\setlength{\extrarowheight}{3pt}
	\centering
		\begin{tabular}{r|ccc|ccc|ccc|ccc|}
	
	 & ~~ &unpoled& ~~ & ~~ &-10~V& ~~ & ~~ &+10~V& ~~ & ~~ &unpoled& ~~ \\
	\hline
	\hline
			Amplitude &\cellcolor[RGB]{0,114,190}&$0.16\pm0.06$& &\cellcolor[RGB]{218,83,25}&$0.16\pm0.07$& &\cellcolor[RGB]{238,178,32}&$0.27\pm0.10$& &\cellcolor[RGB]{126,47,142}&$0.15\pm0.06$&\\ 
			\cline{2-13}
			(arb.units) &\cellcolor[RGB]{119,173,48}&$0.29\pm0.11$& &\cellcolor[RGB]{77,191,239}&$0.48\pm0.13$& &\cellcolor[RGB]{163,20,47}&$0.62\pm0.15$& &\cellcolor[RGB]{214,20,98}&$0.21\pm0.08$&    \\ 
		\hline \hline
			Phase &\cellcolor[RGB]{0,114,190}&$2\pm80$& &\cellcolor[RGB]{218,83,25}&$2\pm86$& &\cellcolor[RGB]{238,178,32}&$-4\pm29$& &\cellcolor[RGB]{126,47,142}&$-2\pm78$ & \\ 
			\cline{2-13}
			(deg.)		&\cellcolor[RGB]{119,173,48}&$-34\pm109$& &\cellcolor[RGB]{77,191,239}&$-65\pm111$& &\cellcolor[RGB]{163,20,47}&$7\pm12$& &\cellcolor[RGB]{214,20,98}&$-25\pm99$ &  \\ 
		\hline \hline
			Lateral Deflection &\cellcolor[RGB]{0,114,190}&$0.45\pm0.02$& &\cellcolor[RGB]{218,83,25}&$0.44\pm0.03$& &\cellcolor[RGB]{238,178,32}&$0.44\pm0.02$& &\cellcolor[RGB]{126,47,142}&$0.47\pm0.02$ &\\ 	\cline{2-13}
			(nm)							 &\cellcolor[RGB]{119,173,48}&$0.38\pm0.02$& &\cellcolor[RGB]{77,191,239}&$0.34\pm0.03$& &\cellcolor[RGB]{163,20,47}&$0.35\pm0.02$& &\cellcolor[RGB]{214,20,98}&$0.40\pm0.02$ & \\ 
		\hline
		\end{tabular}
	\caption{Table with the average values and corresponding standard deviation for the line profile areas in Fig.~\ref{fig:pfm}(e,g,j).}
	\label{tab:pfm}
\end{table}


\section{Topography of the $\mbox{SrTiO}_3$ and $\mbox{KTaO}_3$ substrates}

\begin{figure*}[htbp]
	\includegraphics[width=1\textwidth]{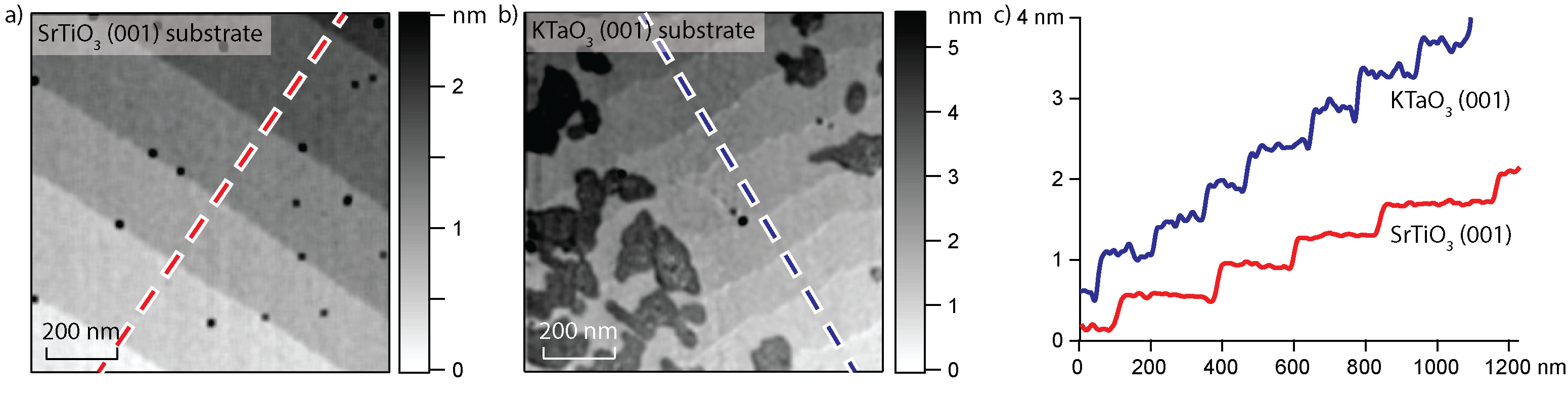}
	\caption{(a) AFM topography of a SrTiO$_3$(001) substrate. (b) AFM topography of a KTaO$_3$(001) substrate. (c) Extracted line profiles along the dashed lines marked in (a,b) averaged over 25~nm width.}
	\label{fig:afm_sub}
\end{figure*}

The atomic force microscopy (AFM) topographies of SrTiO$_3$(001) (STO)  and KTaO$_3$(001) (KTO) substrates as used for the growth of the BTO films of this study, are presented in Fig.~\ref{fig:afm_sub}. These data are taken with a different device than the PFM data in Fig.\ref{fig:pfm}. The STO substrate is etched to obtain a TiO$_2$ terminated surface. The etching procedure is described in the SOM of Plumb et al. \cite{Plumb:2014}. The measured AFM topography in Fig.~\ref{fig:afm_sub}(a) shows the presence of terraces with a width of approximately 210~nm. The KTO substrate is not etched and accordingly shows a mixed termination in the AFM data [Fig.~\ref{fig:afm_sub}(b)] in the form of higher (dark) patches. The observed terrace width is with approximately 120~nm roughly half the observed size of the STO substrate [Fig.~\ref{fig:afm_sub}(c)].

\newpage
\section{Estimation of the Electric Field}

The external electric field of a ferroelectric material is given by its polarization $P$ \cite{FeynmanV2:1964, Hofmann:2014}. For bulk, tetragonal BTO the polarization is reported to be $P\approx0.25~C/m^2$ \cite{Merz:1953} which results in and external electric field of:
\begin{equation}
 |F|=\frac{P}{(\epsilon_r-1)~\epsilon_0}\approx5\cdot10^{8}~\mbox{V/m} 
 \label{eq:efield3}
\end{equation}

Based on this external electric field $F$, the local electric field $F_{loc}$ inside the material is given as $F_{loc}=1/3~(\epsilon_r+2)~F$ \cite{Hofmann:2014}. With the relation \ref{eq:efield3} for the external electric field, the local electric field can be written as a function of the polarization $P$.
  \begin{equation}
     |F_{loc}|=\frac{1}{3}\frac{P\cdot(\epsilon_r+2)}{(\epsilon_r-1)\cdot\epsilon_0}\approx\frac{1}{3}\frac{P}{\epsilon_0}\approx1\cdot10^{10}~\mbox{V/m}
		\label{eq:efield}
  \end{equation}
	
The obtained local electric field is considered relevant for the occurrence of WSL. With a magnitude of $|F_{loc}|\approx1\cdot10^{10}~\mbox{V/m}$ the local electric field of BTO results in a Bloch oscillation time of $\tau_B=\frac{h}{eFa}\approx1\cdot10^{-15}$~s within the unit cell of BTO ($a\approx4$~\AA). For a relaxation time of $\tau=10^{-14}$~s the condition $\tau>\tau_B$ is therefore satisfied in a unit cell of ferroelectric BTO and the occurrence of Bloch oscillations and Wannier-Stark localization is expected.  

In general, the WSL is accompanied by the formation of a Wannier-Stark ladder, a set of electron states separated in energy and space. In superlattices, where a WSL occurs by the help of an externally applied, tunable electric field, indications of a Wannier-Stark ladder are observed \cite{Mendez:1988, voisin:1988}. 
The energy separation between the steps of the Wannier-Stark ladder is given as $\Delta E=eFa$ \cite{Marder:2010, Hofmann:2014} and expected to be between 0.1-6~eV for the films studied, based on an electric field between $5\cdot10^{8}$~V/m to $1.5\cdot10^{10}$~V/m. With the origin of the electric field in the ferroelectric properties of BTO, the local electric field is not expected to be constant due to the variable domain configurations and sizes. Therefore the energy steps of the resulting Wannier-Stark ladder are not isotropic but will vary within the probed area. Therefore, ARPES is not the method of choice to observe these ladders due to the limited coherence length of these states. Local probe techniques using tunneling or optical spectroscopy should be able to address this aspect in future work.

\section{Supplemental ARPES Measurements on BTO}

\subsection{Constant energy surfaces at higher binding energies}

\begin{figure*}[htbp]
	\includegraphics[width=1\textwidth]{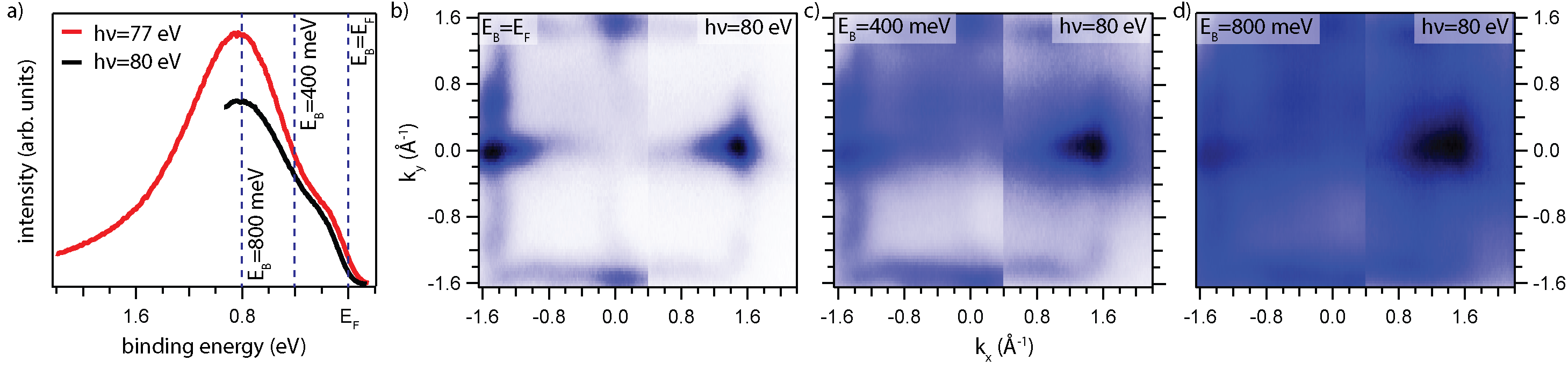}
	\caption{(a) Energy distribution curves for $h\nu=77$~eV and $h\nu=80$~eV at $\overline{\Gamma_{10}}$. (b) Constant energy surface at the Fermi energy, (c) a binding energy of $E_B=400$~meV and (d) a binding energy of $E_B=800$~meV which corresponds to the energy of the in-gap state.}
	\label{fig:cec_ingap}
\end{figure*}

Further constant energy surfaces at higher binding energies, extracted from the same data set shown in Fig.~2(a-c) of the main text, are displayed in Fig.~\ref{fig:cec_ingap}. In Fig.~\ref{fig:cec_ingap} (b), taken at the Fermi energy we clearly see a checkerboard pattern, with strong intensity at the $\Gamma$-points and the smeared intensity along the $\overline{\Gamma\mbox{X}}$ direction, origination from the WSL of the 2D states. At 400~meV binding energy, the pattern observed at the Fermi energy is still visible, however, also spectral intensity around the $\overline{\mbox{M}}$-point, away from the $\Gamma$-points and the $\overline{\Gamma\mbox{X}}$ direction, becomes apparent. At a binding energy of 800~meV, corresponding to the energy of the in-gap state of BTO as seen in Fig.~\ref{fig:cec_ingap}(a) the checkerboard pattern observed at the Fermi energy is not distinguishable anymore. There is a constant intensity background with little structure except higher spectral intensities around the $\Gamma$-points due to diffraction effects. This shows that the checkerboard pattern does not originate from the in-gap states.

\subsection{Measurements along $\overline{\Gamma\mbox{M}}$}

\begin{figure*}[htbp]
	\includegraphics[width=1\textwidth]{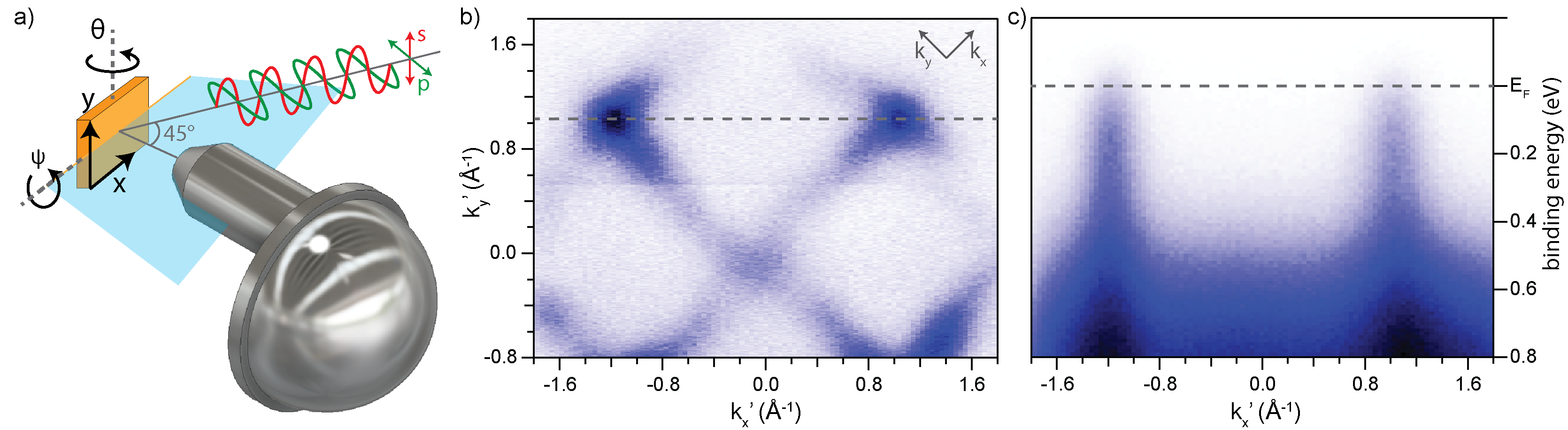}
	\caption{(a) Measurement geometry of the ARPES setup. (b) Fermi surface of 10 uc BTO on STO for $h\nu=77$~eV oriented with $\overline{\Gamma \mbox{M}}$ along the analyzer entrance slit. (c) Band dispersion across $\overline{\Gamma_{10}}$ and $\overline{\Gamma_{01}}$ as indicated in (a).}
	\label{fig:arpes_gm}
\end{figure*}

The angle-resolved photo electron spectroscopy (ARPES) data presented in the main text are all measured with the same geometry where the entrance slit of the hemispherical analyzer is aligned along the $\overline{\Gamma\mbox{X}}$-direction of the crystal. In the measurements in Fig.~\ref{fig:arpes_gm} the crystal is aligned with $\overline{\Gamma\mbox{M}}$ parallel to the analyzer entrance slit [along $\theta$ see Fig.~\ref{fig:arpes_gm}(a)]. The angular scanning direction [$\psi$ in Fig.~\ref{fig:arpes_gm}(a)] is perpendicular to the alignment direction and consequently different for the two cases. Apart from changes in the relative intensities, the altering of the measurement geometry does not affect the data. In particular the WSL states are still visible, smeared along the $\overline{\Gamma\mbox{X}}$-direction. This confirms that our observations are not measurements artifact caused by the probing geometry.
 
The data of Fig.~\ref{fig:arpes_gm} are taken from a sample of 10 uc BTO deposited on a SrTiO$_3$ (STO) substrate. In comparison with the data on 20 uc BTO on STO presented in the Fig.3 of the main text, the smearing tend to be more uniform. Possible reasons for this are the combined effect of altered matrix element contributions due to the different measurement geometry and a different domain pattern as a consequence of the reduced film thickness.

\subsection{Photon Energy Dependency}

\begin{figure*}[htbp]
	\includegraphics[width=1\textwidth]{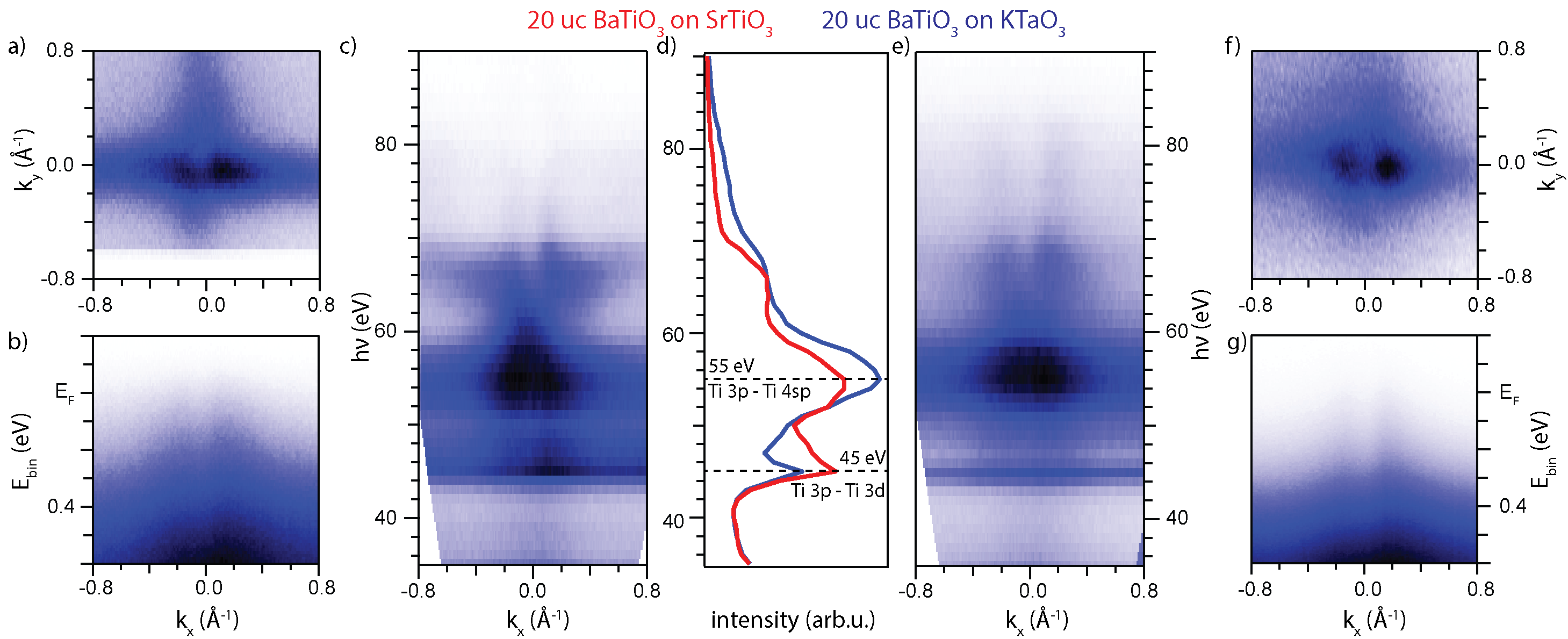}
	\caption{ARPES data of 20 uc BTO grown on STO (a-c) and on KTO (e-g). (a,f) Fermi surface at $h\nu=67$~eV and (b,g) Band dispersion in $\overline{\Gamma \mbox{X}}$-direction. (c,e) Photon energy scan from $h\nu=35$~eV to $90$~eV at the Fermi energy. (d) Integrated photoemission intensity over $k_x=\pm 0.6$~\AA$^{-1}$ at the Fermi energy for the sample grown on STO (red) and KTO (blue) substrate.}
	\label{fig:arpes67eV}
\end{figure*}

Fig.~\ref{fig:arpes67eV} depicts measured Fermi surfaces, band maps, and photon energy dependence of the two-dimensional states around $k_x=0$~\AA$^{-1}$, the $\overline{\Gamma_{00}}$ point. The measured intensity at the Fermi energy as a function of $k_x$ and photon energy [Fig. \ref{fig:arpes67eV}(c,e)] shows bands, forming two parallel lines with photon energy close to $k_x=0$~\AA$^{-1}$, that corresponds to $\overline{\Gamma_{00}}$. The different photon energies give access to different $k_z$. Therefore the lack of dispersion of these two parallel bands with photon energy indicates their two-dimensional (or one-dimensional) nature.
The observed intensity modulation with photonenergy is given by the excitation to different available final states as well as resonant enhancements. The photon energies of $h\nu=45$~eV and $h\nu=55$~eV [Fig.~\ref{fig:arpes67eV}(d)] correspond to the energies of the Ti 3p - Ti 3d and Ti 3p - Ti 4sp resonance, respectively \cite{Smith:1988, Tao:2011}. The Ti 3p - 3d resonance has a sharp Fano-like lineshape \cite{Fano:1961} indicating a low dimensionality of the excited state \cite{Tao:2011}; i.e. the Ti 3d states close to the Fermi level. On the other hand, the Ti 3p - Ti 4sp resonance is much broader implying that these states, hybridized with oxygen, are more delocalized along the z-direction \cite{Tao:2011}.

\subsection{Light Polarizations}

\begin{figure*}[b]
	\includegraphics[width=0.8\textwidth]{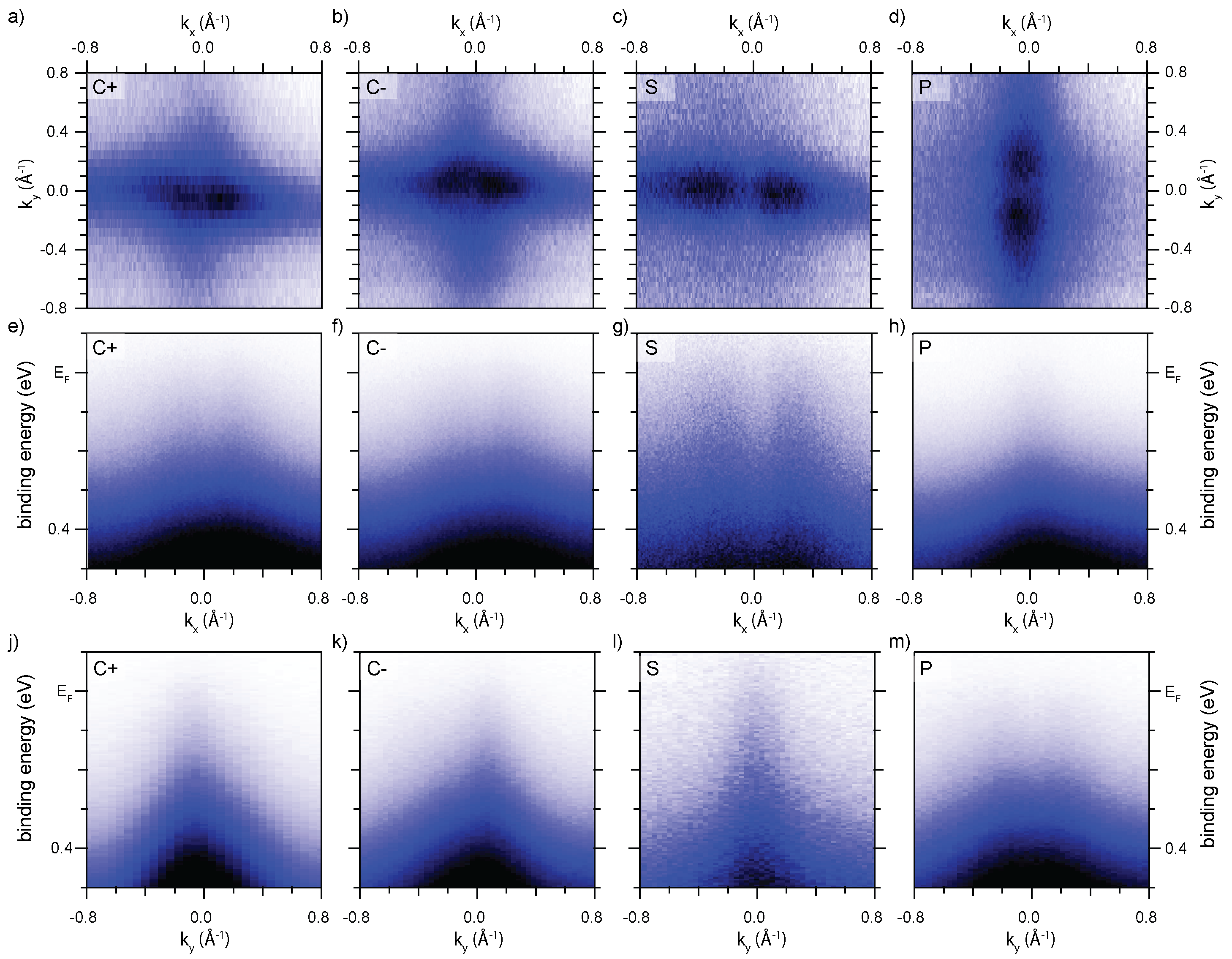}
	\caption{ARPES data of 20 uc BTO on STO for different light polarizations measured with $h\nu=67$~eV. (a-d) Fermi surfaces of the first Brillouin zone around $\overline{\Gamma_{00}}$ for (a) left- and (b) right-hand circularly polarized light and linear (c) s- and (d) p-polarized light. (e-h) Band dispersions along the $k_x$-direction for $k_y=0$~\AA$^{-1}$ and (j-m) band dispersions along the $k_y$-direction ($k_x=0$~\AA$^{-1}$) for the four light polarizations.}
	\label{fig:pol_g0}
\end{figure*}

\begin{figure*}[htbp]
	\includegraphics[width=0.8\textwidth]{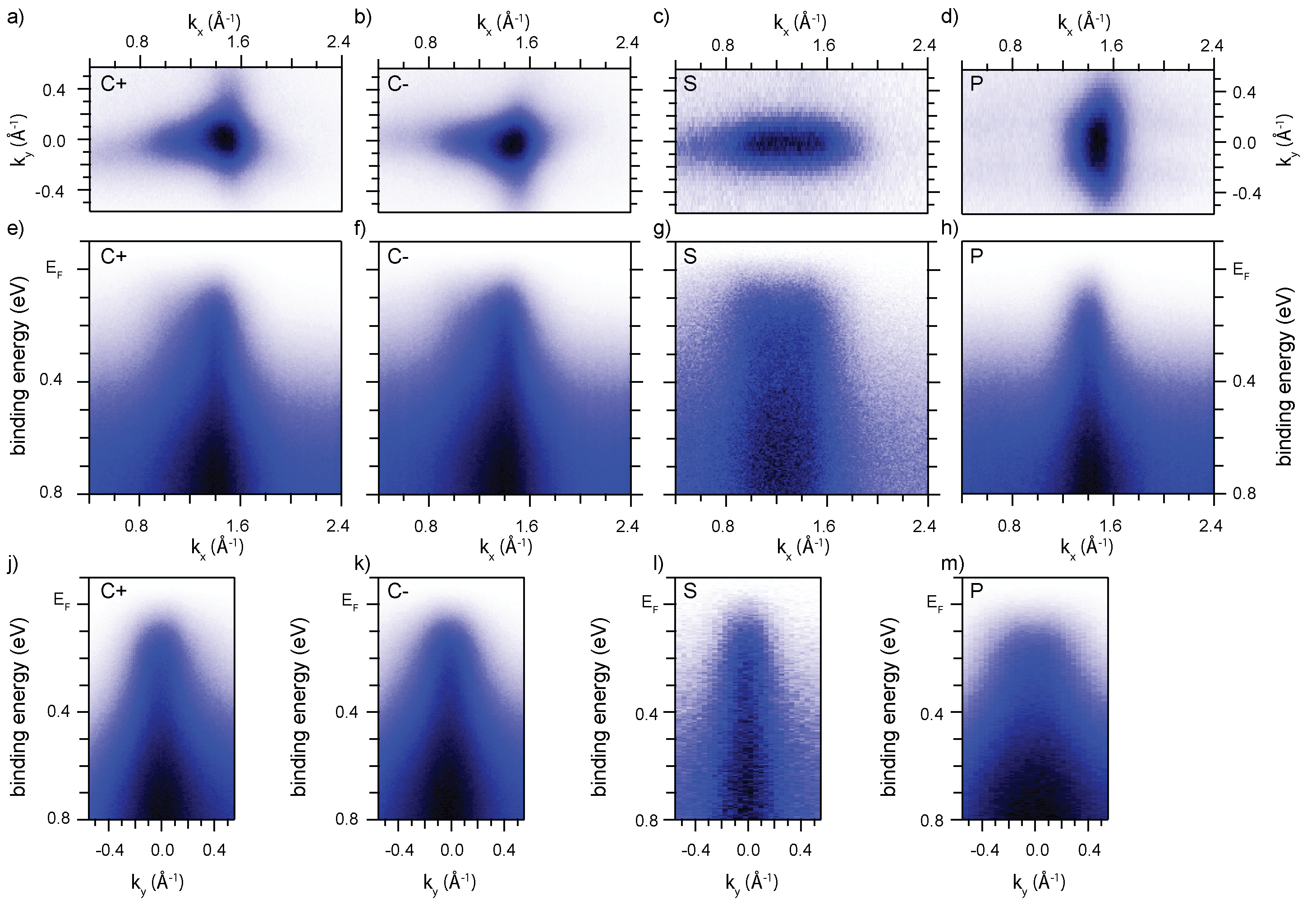}
	\caption{Light polarization dependent ARPES of 20 uc BTO on STO measured with $h\nu=80$~eV. (a-d) Fermi surfaces of the second Brillouin zone showing $\overline{\Gamma_{10}}$ for (a) C+, (b) C-, (c) s- and (d) p-polarized light. (e-h) Corresponding band dispersions along the $k_x$-direction for $k_y=0$~\AA$^{-1}$ and (j-m) band dispersion along the $k_y$-direction at $k_x=1.57$~\AA$^{-1}$ for the four light polarizations. }
	\label{fig:pol_g2}
\end{figure*}

The Fermi surfaces and corresponding band structures for right- and left hand circularly polarized light as well as for s- and p-polarized linear light are depicted in Fig.~\ref{fig:pol_g0} and Fig.~\ref{fig:pol_g2}. For these data, the analyzer entrance slit is aligned along the $\overline{\Gamma \mbox{X}}$-direction of the sample as in the main text. For s-polarized light the electric field of the synchrotron light is along the $k_y$-direction, for p-polarized light along the $k_x$-direction [see Fig.~\ref{fig:arpes_gm}(a)].
In the data in the vicinity $\overline{\Gamma_{00}}$ [Fig.~\ref{fig:pol_g0}] measured with a photon energy of $h\nu=67$~eV no differences are noticeable for the two circular polarizations [Fig.~\ref{fig:pol_g0}(a,b)]. The band dispersion along the $k_x$-direction for circularly polarized light shows two features, connected to the in-gap state as discussed in the main text. Along $k_y$ the band structure only hosts a single intensity feature at $k_y=0$~\AA$^{-1}$. 

For s-polarized light [Fig.~\ref{fig:pol_g0}(c)], the Fermi surface consist of two features elongated along the $k_x$-direction, separated by suppressed intensity at $k_x=0$~\AA$^{-1}$. Accordingly two features appear in the $k_x$ dispersion, and the $k_y$ dispersion only shows enhanced intensity around $k_y=0$~\AA$^{-1}$. The Fermi surface with p-polarized light [Fig.~\ref{fig:pol_g0}(d)] is similar to the one measured with s-polarized light but $90^{\circ}$ rotated, with suppressed intensity along $k_y=0$~\AA$^{-1}$.

The suppressed intensity for $k_x=0$~\AA$^{-1}$ with s-polarized light and for $k_y=0$~\AA$^{-1}$ with p-polarized light is an indication for an xy-symmetry of the probed orbitals. In the case of the other known perovskites hosting a two-dimensional electron gas \cite{Santander:2011, Meevasana:2011, King:2012, Santander:2012, Plumb:2014, Muff:2017}, the two-dimensional states are attributed to the Ti~3$d_{xy}$ orbitals and the three-dimensional bands, dispersing with photon energy, with the Ti 3$d_{xz}$ and Ti 3$d_{yz}$ orbitals. It seems likely for BTO to have a similar orbital ordering. However, due to the WSL of the states at the BTO surface, the orbital symmetries of the states present cannot be conclusively assigned. 

The data in Fig.~\ref{fig:pol_g2} of the Fermi surface at $\overline{\Gamma_{10}}$ for the different light polarizations are very similar to the data of $\overline{\Gamma_{00}}$ with respect to the light polarization effects. 
For right- as well as left-hand circularly polarized light [Fig.~\ref{fig:pol_g2}(a,b)] the intensity of the WSL along $\overline{\Gamma\mbox{X}}$ is visible. However, the intensity distribution along $k_y$ around the $\Gamma$-point is inverted. For the linear polarized light only the WSL states in $k_x$-direction are visible for s-polarized light, while for p-polarized light only intensity elongated in $k_y$-direction are present. 
In contrast to the data taken at $\overline{\Gamma_{00}}$, the suppresion of intensity at $k_x=0$~\AA$^{-1}$ or $k_y=0$~\AA$^{-1}$ for linear polarized light is absent due to the emission angle of $\overline{\Gamma_{10}}$ being far off normal emission. The band dispersion along the $k_y$-direction for the four different light polarizations consist of a main feature around $\overline{\Gamma_{10}}$, dispersing from the in-gap state. The band dispersions for the circular polarized lights, show an asymmetry around $k_y=0$~\AA$^{-1}$ corresponding with the Fermi surface. For s-polarized light the feature is narrow in the $k_y$-direction while for p-polarized light it is broad, opposite to the band dispersions along the $k_x$-direction.

\newpage
\subsection{Time Dependent Behavior}
\label{sec:bto_time}

The BTO films show a time dependent behaviour under UV-irradiation. In order to turn the surface conductive and avoid charging, a path is written by the UV-light starting from the mounting clamp to the center of the sample. This is an established experimental procedure for ARPES measurements of the 2D states of titanates surfaces (see SOM \cite{Plumb:2014}). 

\begin{figure*}[htbp]
	\includegraphics[width=0.75\textwidth]{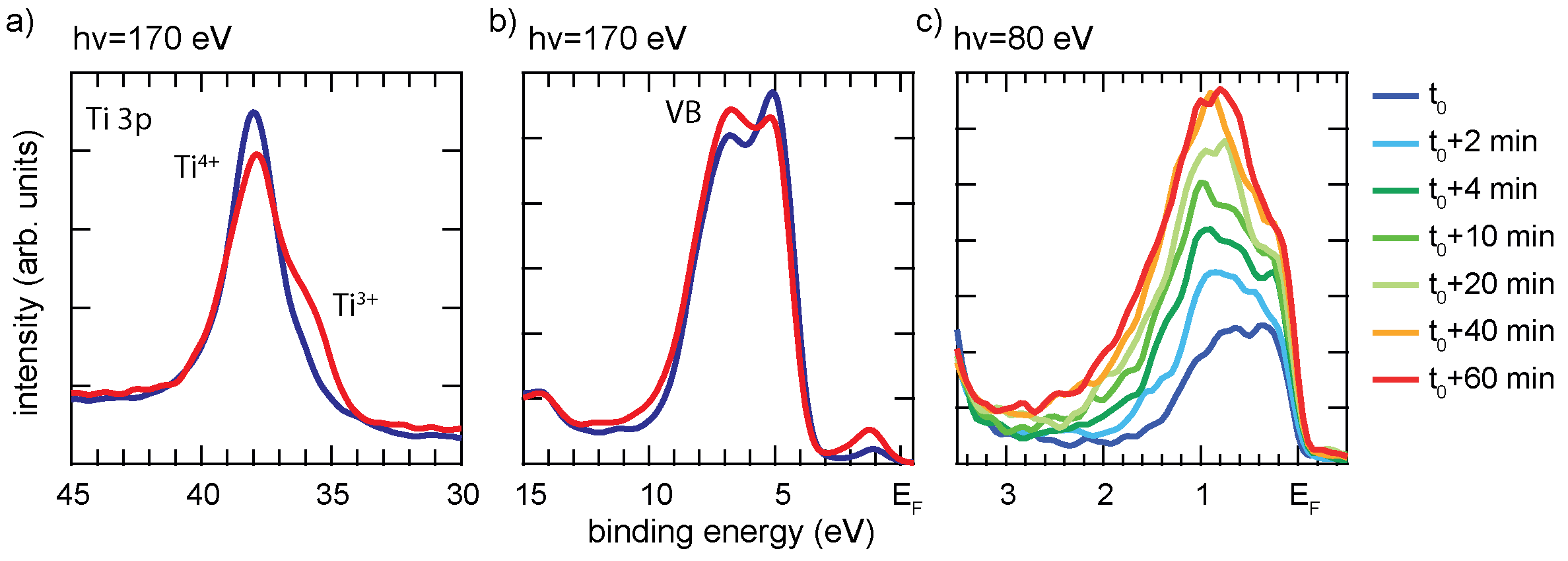}
	\caption{Time-dependent, angle-integrated photoemission intensity of the (a) Ti 3p core level (b) valence band, measured with $h\nu=170$~eV and (c) in-gap and surface states, measured with $h\nu=80$~eV.}
	\label{fig:time}
\end{figure*}

Under irradiation, the spectral intensity of the in-gap state at 0.8~eV binding energy and of the Ti~$3^+$ shoulder of the Ti 3p core level is increasing with time [see Fig.~\ref{fig:time}]. In case of the Ti~$3^+$, its percentage on the total Ti~3p peak area rises from 7\% to 18\% within one hour [Fig.~\ref{fig:time}(a)]. This scales to a free charge carrier density at the surface of 0.18 electrons per uc after one hour, when saturation is reached. Within the same time frame, the intensity of the in-gap state rises by 300\% [Fig.~\ref{fig:time}(b,c)]. 
However, the metallic state, visible as a second peak at the Fermi energy [Fig.~\ref{fig:time}(c)], only changes by 0.3\% in peak area within the same time frame. Thus although the intensity of Ti~$3^+$ and the in-gap state seem to be related, the intensity of the metallic states does not directly scale.
The surface localized Ti~$3^+$ ions are linked to the creation of oxygen vacancies and structural reordering of the surface layers in the titanium based perovskites \cite{Plumb:2014}. The changes implied in the distortion of the TiO$_6$ octahedra and their respective binding angles due to the reordering will alter the hybridization of titanium and oxygen. Indications of this change in hybridization are observable in the altering peak intensity of the valence band with time.   
The observed changes under UV-light saturate within 30~min and are persistent with time regardless if the area is further irradiated or not.


\section{Further Characterization of STO thin Films}

\subsection{RHEED pattern and Oscillations}

\begin{figure*}[htbp]
	\includegraphics[width=0.7\textwidth]{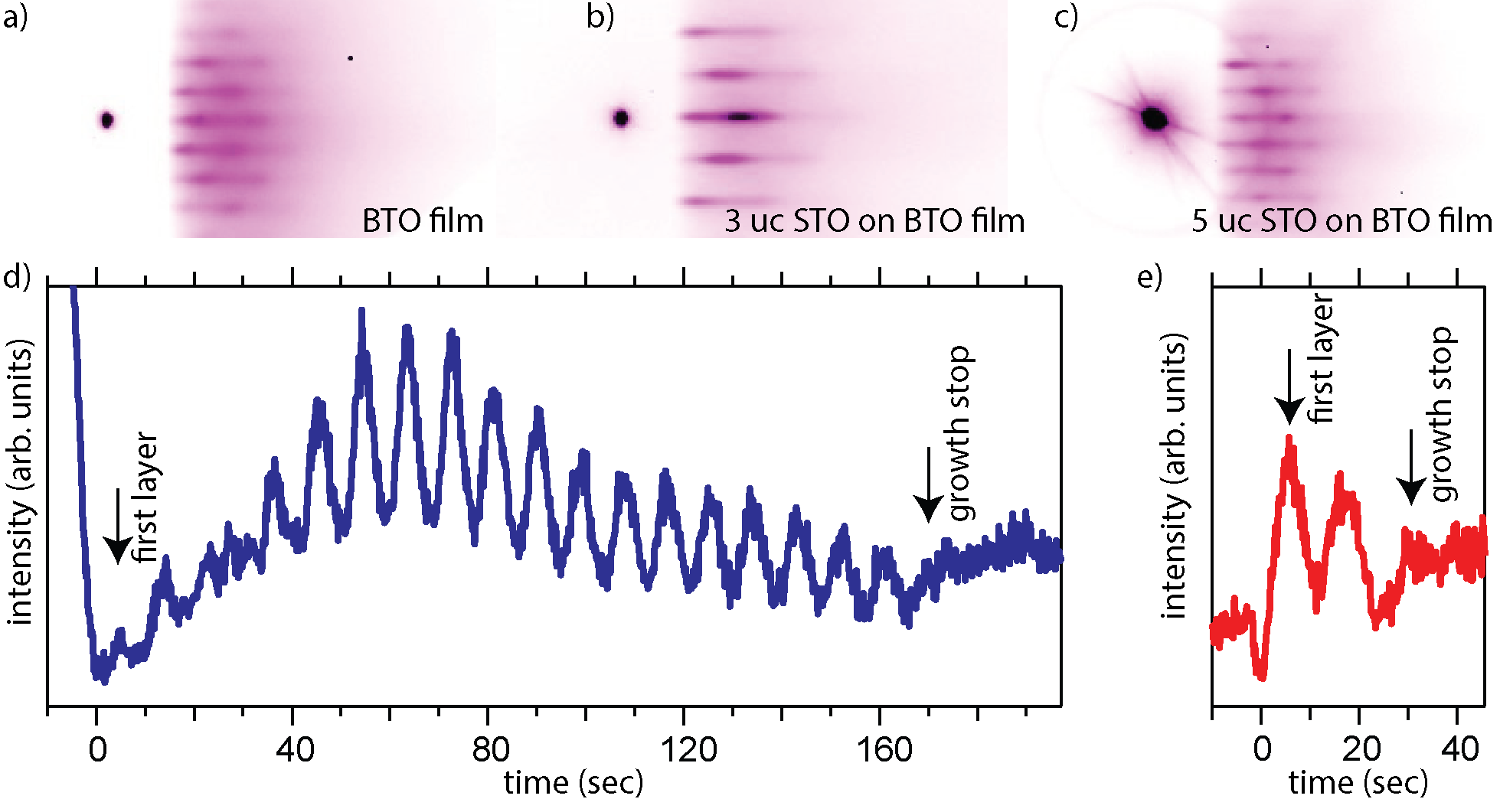}
	\caption{(a) RHEED pattern of 20 uc BTO film grown on STO , (b) pattern of 3 uc STO grown on a BTO film and (c) pattern of 5 uc STO grown on a BTO film. (d) RHEED oscillations during the growth of the BTO film. (e) RHEED oscillations of 3 uc STO grown on a 20 uc STO film.}
	\label{fig:supl_rheed}
\end{figure*}

The growth process with pulsed laser deposition (PLD), was monitored by reflective high-energy electron diffraction (RHEED) patterns and oscillations. The RHEED pattern and oscillations of a film of 20 uc BTO grown on a STO substrate are depicted in Fig.~\ref{fig:supl_rheed}(a,d). The RHEED pattern was obtained after the growth and indicates a crystalline two-dimensional surface. Each maxima of the RHEED oscillation corresponds to the formation of a complete BTO layer and therefore allows a precise thickness control of the film while growing. 
The RHEED pattern of the 3 uc [Fig.~\ref{fig:supl_rheed}(b)] and 5 uc [Fig.~\ref{fig:supl_rheed}(c)] film of STO deposited on a previously grown BTO film of 20 uc shows a good crystalline surface. By the help of the RHEED oscillations of the STO thin film growth [Fig.~\ref{fig:supl_rheed}(e)], a precise termination of the growth process is possible at the oscillation maxima.  
 
\newpage
\subsection{XPS Measurements}

\begin{figure*}[htbp]
	\includegraphics[width=0.8\textwidth]{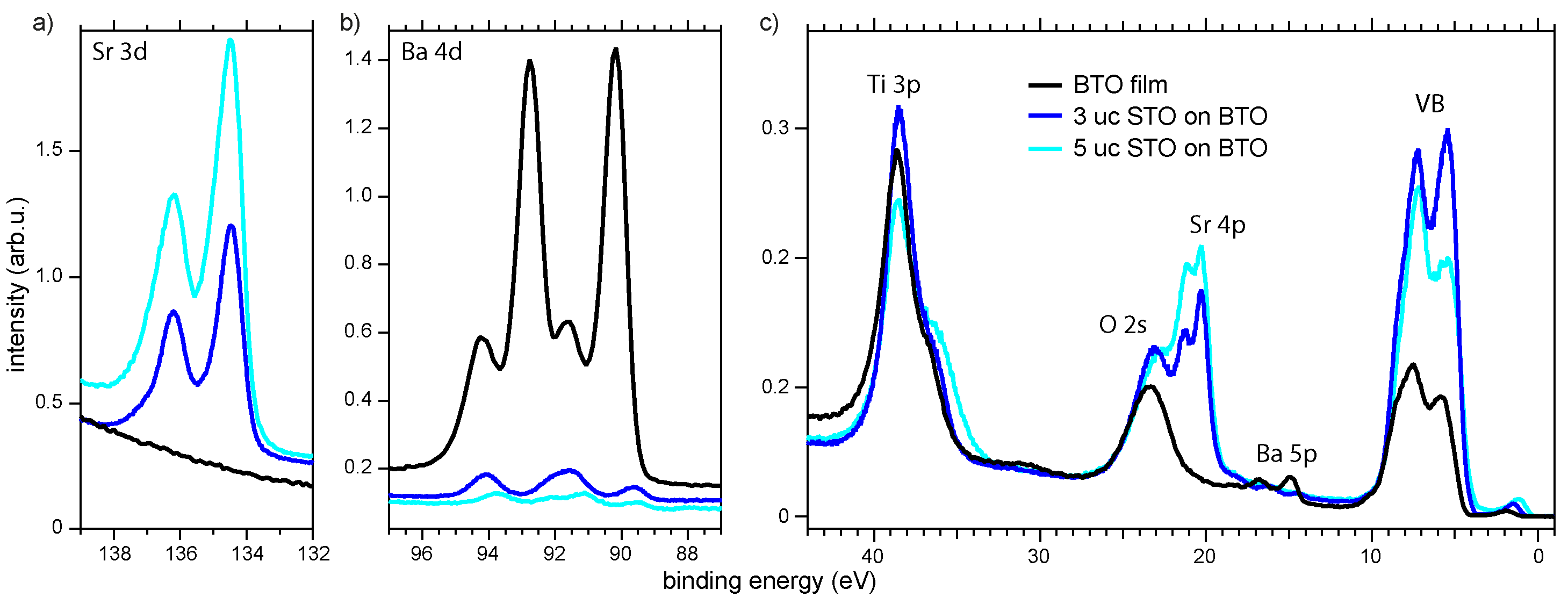}
	\caption{XPS spectra obtained with $h\nu=170$ eV for the clean BTO film and with 3 and 5~uc of STO on top. (a) Sr 3d, (b) Ba 4d, (c) Ti 3p, O 2s, Sr 4p, Ba 5p, and the valence band.}
	\label{fig:xps_comp}
\end{figure*}

In Fig.~\ref{fig:xps_comp} a comparison of XPS spectra is shown for the clean BTO films and 3 and 5~uc of STO grown on top. The data were normalised to the background after the O 2s core level and the BTO data was offset in Fig.~\ref{fig:xps_comp}(a) for clarity. As expected the Sr core levels increase with STO coverage whereas the Ba core levels show an exponential decay with coverage and are almost completely suppressed for the 5~uc thick STO film. This indicates a layer-by-layer growth of a closed STO film on top of the BTO substrate.

The small changes of binding energies in the Ba 4d core levels could give insight in the detailed atomic structure of the BTO/STO interface and possible intermixing in the first unit cell. However, this goes far beyond the scope of this work and is best combined with detailed structural investigations.

\newpage
\subsection{ARPES Measurements}

\begin{figure*}[htbp]
	\includegraphics[width=0.8\textwidth]{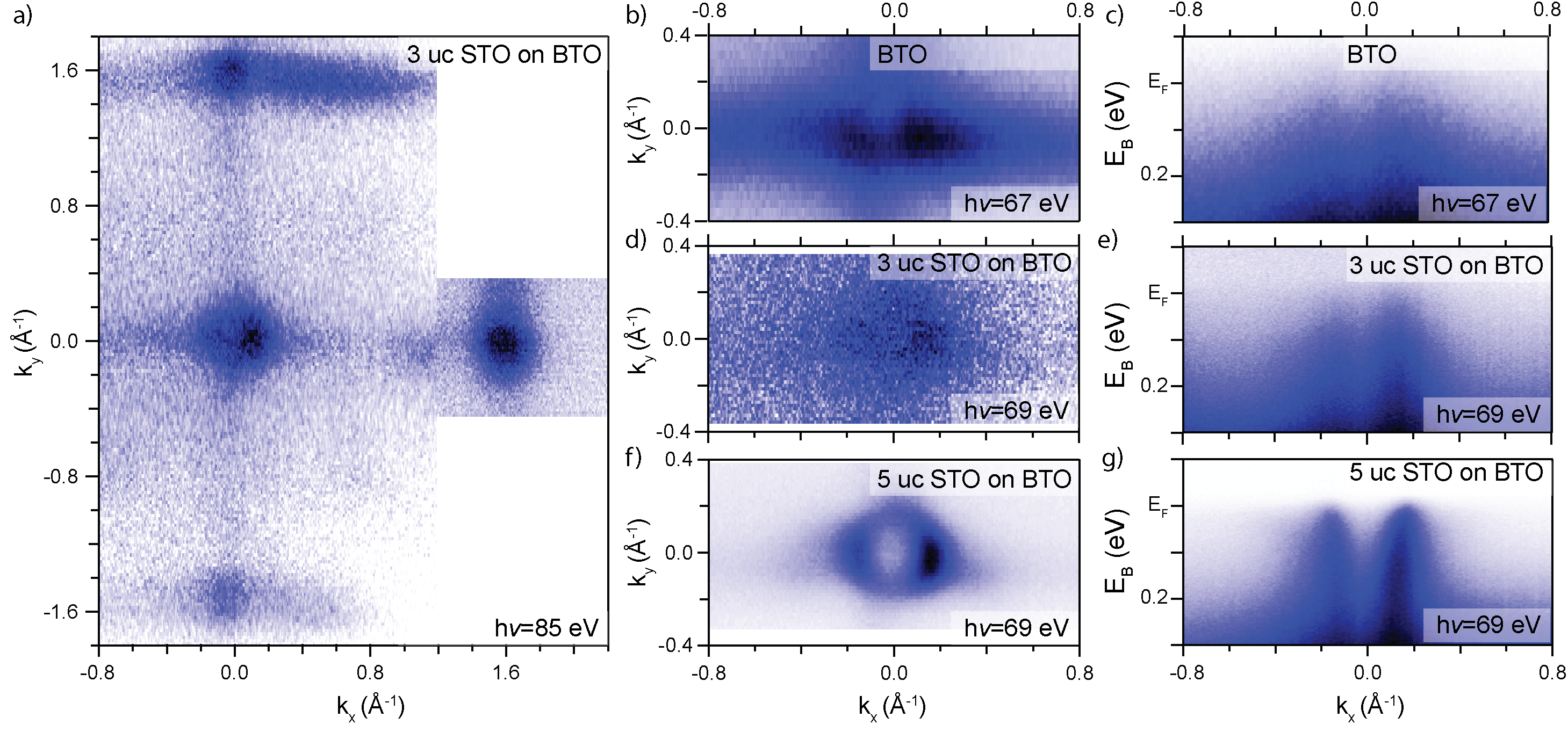}
	\caption{(a) Fermi surface of 3~uc of STO on BTO obtained at $h\nu=85$~eV. (b,d,f) Fermi surfaces around the surface Brillouin zone center for pure BTO, 3~uc, and 5~uc of STO on BTO at the indicated photon energies. (c,e,g) Band maps at $k_y=0$ for  pure BTO, 3~uc, and 5~uc of STO on BTO at the indicated photon energies.}
	\label{fig:arpes_sto}
\end{figure*}

In Fig. 3(a) of the main text a subsection of the Fermi surface for a 3~uc thick film of STO grown on a 20~uc film of BTO was shown. Although the WSL states are readily discernible their extension becomes more clearly visible in the large range Fermi surface map in Fig.~\ref{fig:arpes_sto}(a).

In Fig~\ref{fig:arpes_sto}(b-g) a comparison between pure BTO, 3~uc of STO on BTO, and 5~uc of STO on BTO is shown for a different photon energy as in Fig. 3 of the main text. Although the intensity ratio between the 3$d_{xy}$ and 3$d_{xz}$ (or 3$d_{yz}$) bands has changed for the 5~uc data, the general features are independent of photon energy.

\bibliographystyle{apsrev4-1}

\pagenumbering{gobble}

\end{document}